# COMPACT GANTRY SYSTEM FOR THE IRRADIATION OF A LYING PATIENT BY PROTON/ION BEAM

A.Mikhailichenko, CLASSE, Ithaca, NY 14850

***Abstract.*** We describe here a compact gantry system, when the lying patient is slowly moved in horizontal and vertical directions in accordance with rotation of transport channel around its axis so that the region, irradiated from different directions, stays in a focus of beam optical system. By this combination of rotation of magnets and linear motion of patient it becomes possible to simplify magnetic system significantly.

## INTRODUCTION AND BACKGROUND

The traditional system for the irradiation of a patient's tumor with an accelerated ion/proton beam is represented schematically in Fig.1, see for example [1]-[5]. It contains a proton/ion accelerator, the transport channel and the system, called a gantry for simplicity. The gantry system holds the terminal part of a transport channel, and has the ability to spin the elements of this channel around the axis of immovable rest part of the beam channel. This configuration allows irradiation of a desirable point in the patient's body under different angles. The last in turn, allows reduction of accumulated dose in healthy tissue around the tumor. That is the goal of all arrangement. In addition to these simple (geometrical) arrangements, in contrast with irradiation by X- rays (where such arrangement is working also), irradiation by a proton or even ion beam brings additional relief for the neighboring healthy tissue thanks to increased ionization right before the particle comes to rest. The last is a physical factor as the ionization is reversely proportional to the squared velocity of an ionizing particle and drastically increases while the proton/ion slows down by losing its energy in a tissue and comes to rest (Bragg peak), as the stopping power, $dE/dx @ e^2 Z^2 / v^2$. So by manipulating accurately with energy of the beam ejected from the accelerator, one can enhance the contrast ratio of the dose distributed in the tumor and around.

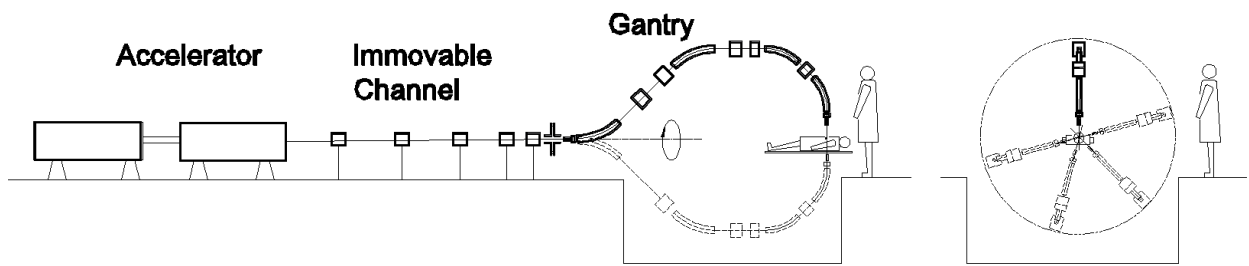

**Figure 1.** Traditional proton/ion complex with gantry system is represented in two orthogonal projections.

In Fig. 2 the typical complex with cyclic accelerator, distribution system, the gantry system which sweeps the beam channel within the cone (marked as 18) is represented in isometric projection [4] as it appears in one of the U.S. patents.

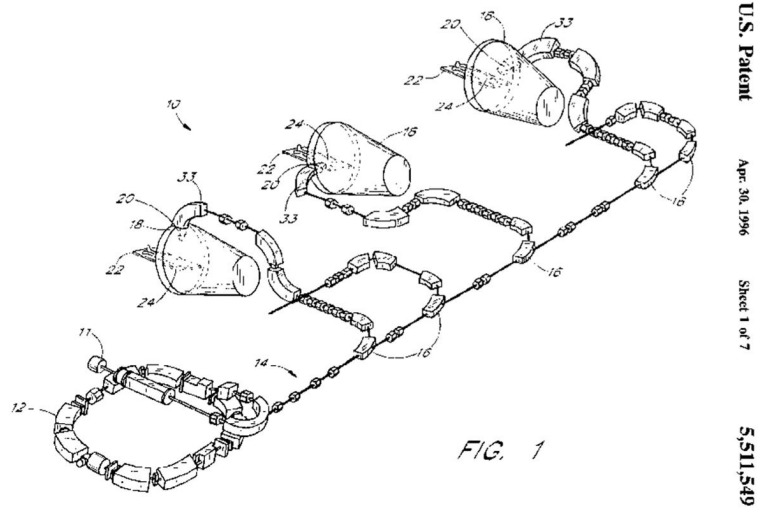

**Figure 2.** Ion/Proton complex [4]. It includes an accelerator, distributional channels, and multiple gantry systems. In this complex, the patient 22 stays immovable at all times, while the end of beam channel spins around.

So one can see, that the system has (two) magnets, bending the beam aside of the transport axis and the magnets which bend the beam to 90° with respect to the immovable transport part of channel. The beam chamber in a region of twist is made hermetically sealed with the ability to twist. There are many ways to do this.

One other approach – to irradiate the sitting or standing patient, which could simplify the system even more –is less recognized due to its less comfortable condition for the patient. However for neutron therapy it is the only way to go as the neutron beam could not be bent easily to the significant angle. Such systems used at FermiLab [6] for neutrons and in [5] for protons.

## THE NEW CONCEPT

The concept of the new system which looks more comfortable, than the one designed for irradiation of sitting or standing patient is represented schematically in Fig.3. In this system the radial distance from the rotation axis to the focal point is kept constant; the patient is lying horizontally on a bed (lodgment), which is moved slowly in horizontal and vertical directions in accordance with position of focal point, so the focal point is irradiated at all times, but under different angles with respect to the patient.

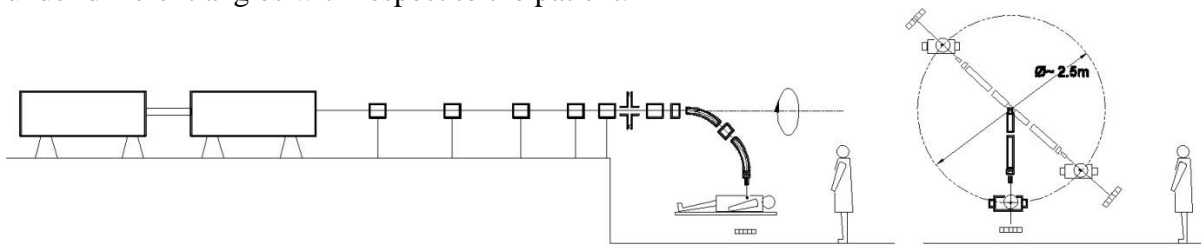

**Figure 3.** New gantry system is represented in two orthogonal projections. Diameter 2.5 m indicated serves just for the reference.

Another peculiarity of the system suggested is that the point of irradiation is the focal point of the beam optics. This focal point is a point with minimal beta-function value. As the envelope function becomes quickly divergent apart from the focal point, $b(s) = b_0 + s^2/b_0$, where $b_0 \sim 1cm$ is a beta function in a crossover (focal point), $s=0$. The beam size is $s(s) = \sqrt{\varepsilon b(s)}$, where $\varepsilon$ stands for emittance of beam. So this fact adds to reduction of dose for the regions around the focal point.

## OPTICS

Beam optics is geometry is represented in Fig.4; all calculation carried for this publication with the OPA code [7].

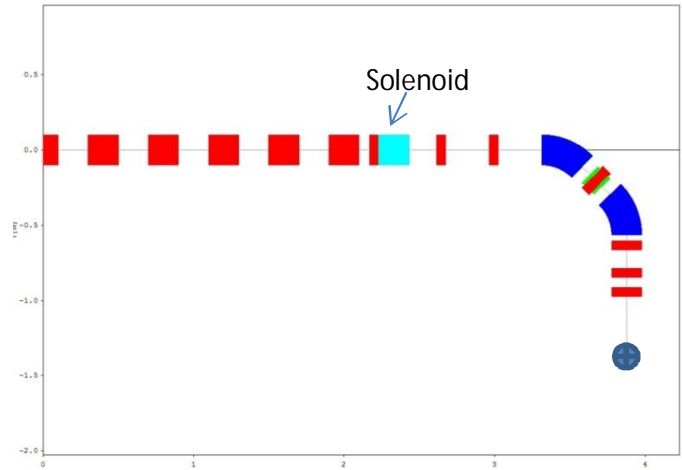

**Figure 4.** Beam optical channel (OPA). Focal point marked by blue cross. Basically this is an 90° achromatic bend.

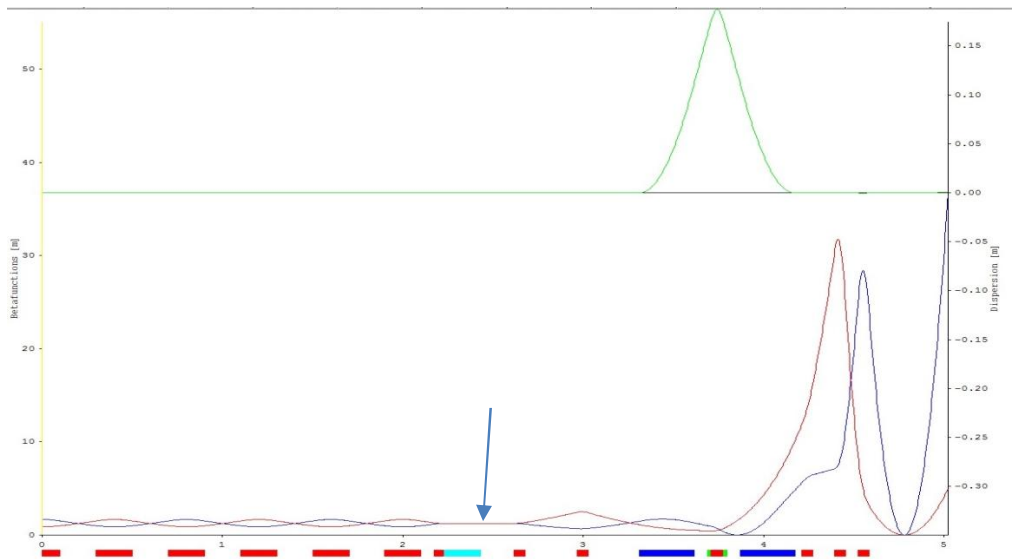

**Figure 5.** Envelope functions for the beam channel from Fig.4. Upper graph represents a dispersion function. The beam optics is achromatic down to the focal point. By arrow marked a place near solenoid, where the envelope function does not change much.

The beam channel after the solenoid is attached to the gantry system by hermetic muff, which allows free spinning along axis. Many of these possible technical solutions could be suggested. The simplest one might be a Wilson type. Further, the part of channel continued with bellow. The envelope functions are the same for both transverse directions. In case if in addition the emittances in transverse plane are the same also, then rotation of channel does not manifest itself at all. As the envelope functions are practically stays horizontal this place could be used as a trombone type joint. This allows longitudinal motion of the elements allocated on gantry frame as whole (kind of a trombone).

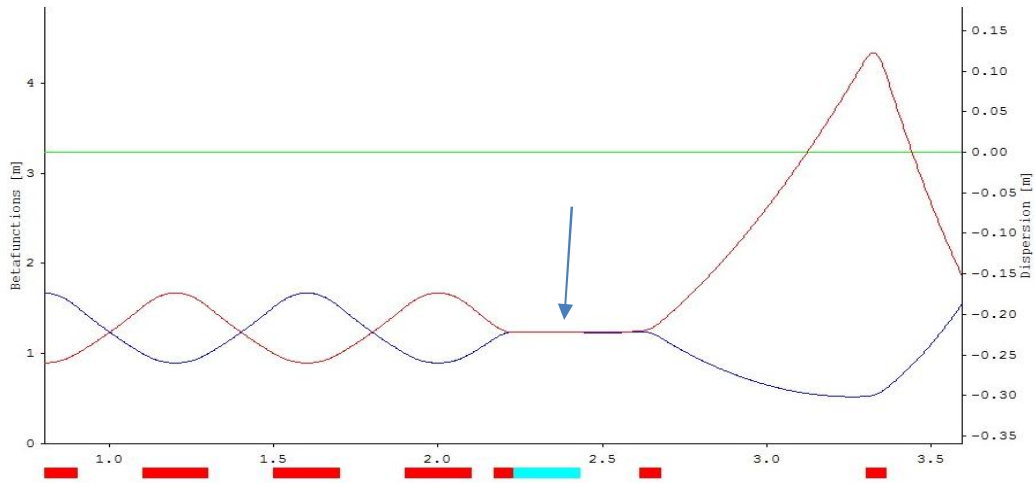

**Figure 6.** Region around rotatable part of channel (marked by arrow in Fig.5, is zoomed.

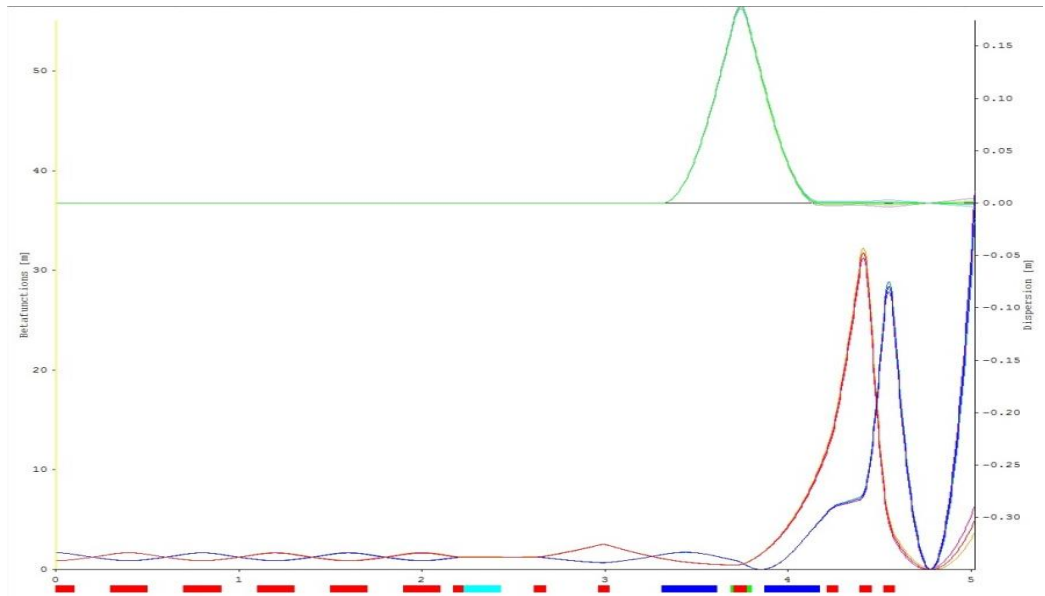

**Figure 7.** Chromaticity of envelope functions and dispersion for $\Delta p/p=\pm 1\%$

Behavior of envelope function in a region marked by arrow in Fig.6 described by the same formula $b(s) = b_0 + s^2 / b_0$, but now the $b_0$ value is of the order of 1.5 $m$, what gives a margin for enlargement of envelope function. So the channel here might be expanded geometrically by the same value (below we suggested movement on $\pm 10$ $cm$).

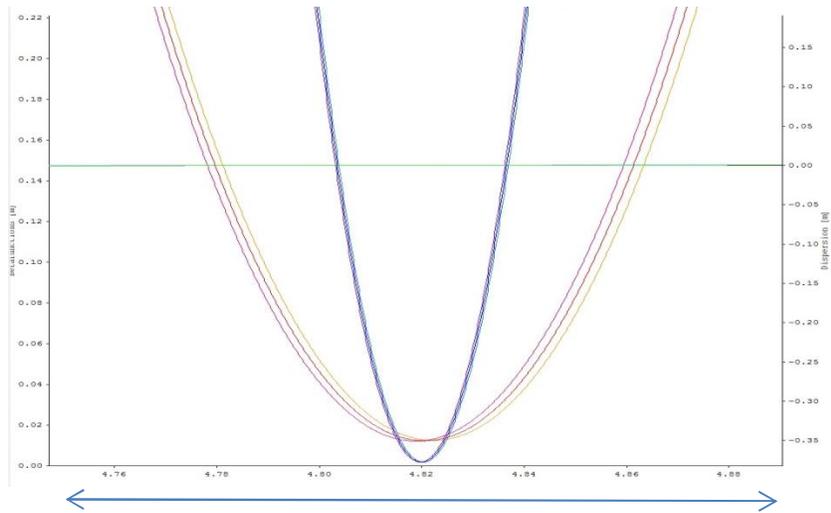

**Figure 8.** Variation of $b_x, b_y$ for momentum deviation $\Delta p/p=\pm 0.1\%$ in a region around the focal point. Longitudinal margins marked by double head arrow on this graph are $\sim$15 *cm*.

In Fig.8 the variation of envelope function is represented, while the momentum of the beam varying by $\Delta p/p=\pm 0.1\%$ . This variation reflects residual chromaticity of beta-functions.

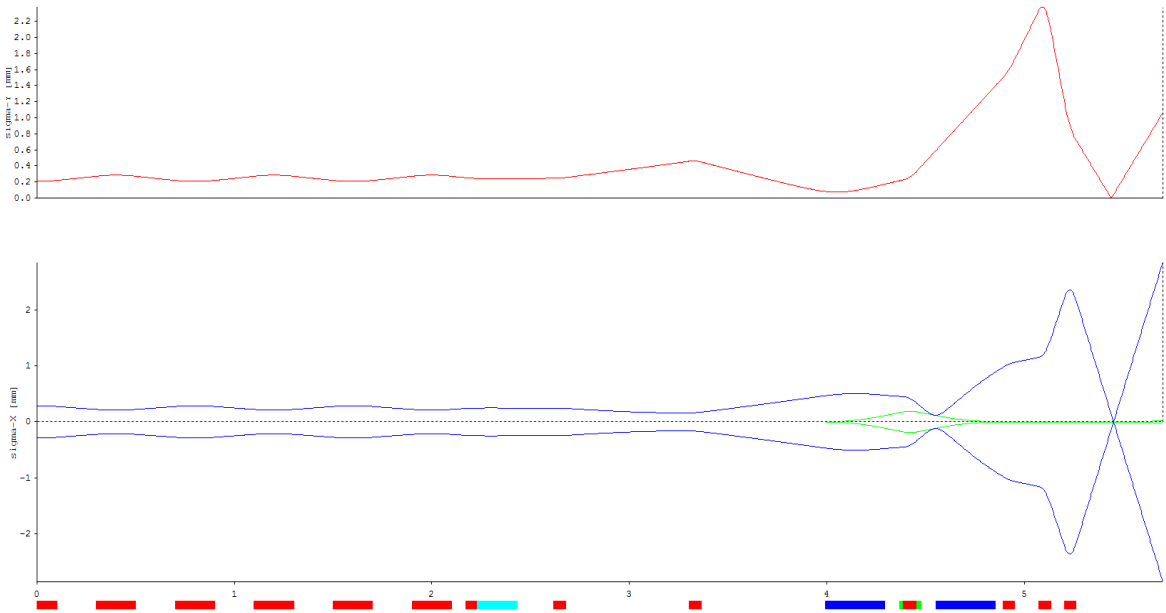

**Figure 9.** The beam envelope for emittance 50 *mm* x *mrad*.

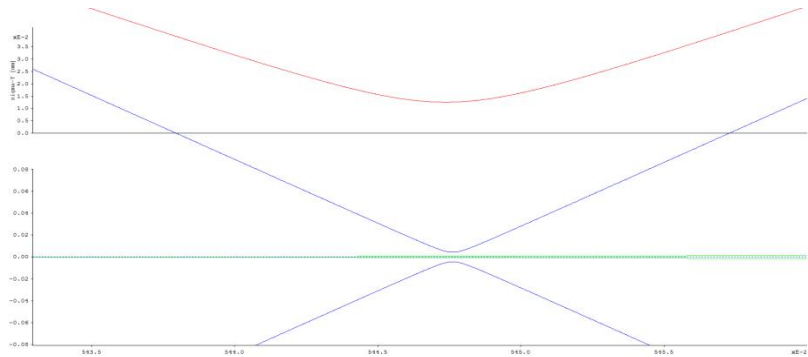

**Figure 10.** Beam size for emittance 50 *mm* x *mrad* zoomed around the focal point. Longitudinal margins of this graph are ~3 *cm*.

One can see from Fig.10 that the beam size in a crossover is 0.02 x 0.005 $mm^2$. One can easily scale these dimensions to any other than 50 *mm* x *mrad* emittance.

By slight variation of the final quadrupole doublet strength, one can move position of waist in longitudinal direction (inside the tumor). A set of dipole correctors allow sweeping of the focal point in transverse directions.

As far as the emittance value, the 50 *mm* x *mrad* is pretty typical. If however, in a future, someone will need even smaller emittances, then the *electron cooling* technics [8] could be implemented here easily. Of cause this will require more space (longer straight sections) and investments, but the good quality beam will pay for this. Extraction of proton/ion beam is going on sum resonance, while the energy of the synchrotron kept constant as required for current position of the beam with respect to the patient's body. Thin electrostatic septum could help in this. Other solution might be associated with sub-picosecond kicker; the pulsed HV generators for feeding such kickers are available on market.

## VARIANT OF EMBODIMENT

One possible embodiment of this new concept represented in Fig.12. Here the gantry holds the beam channel (which is basically a 90° achromatic bend) and a patient also who is placed in a special cradle. This cradle keeps the body of patient in a horizontal position. In addition, the end holders of the cradle have a possibility to adjust the position so the focal point lays on the rotational axis of these holders (see Fig.12). Below the patient table some registration equipment is located. Mechanical holder for this equipment is attaching it to the gantry arc.

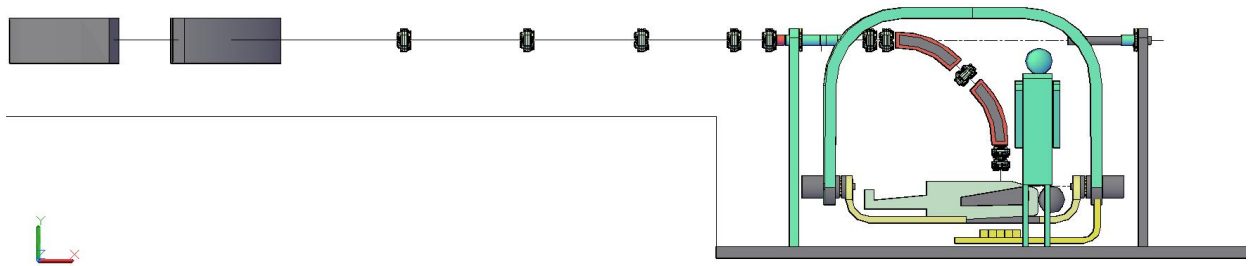

**Figure 11.** Conceptual 3D sketch of system from Fig.3

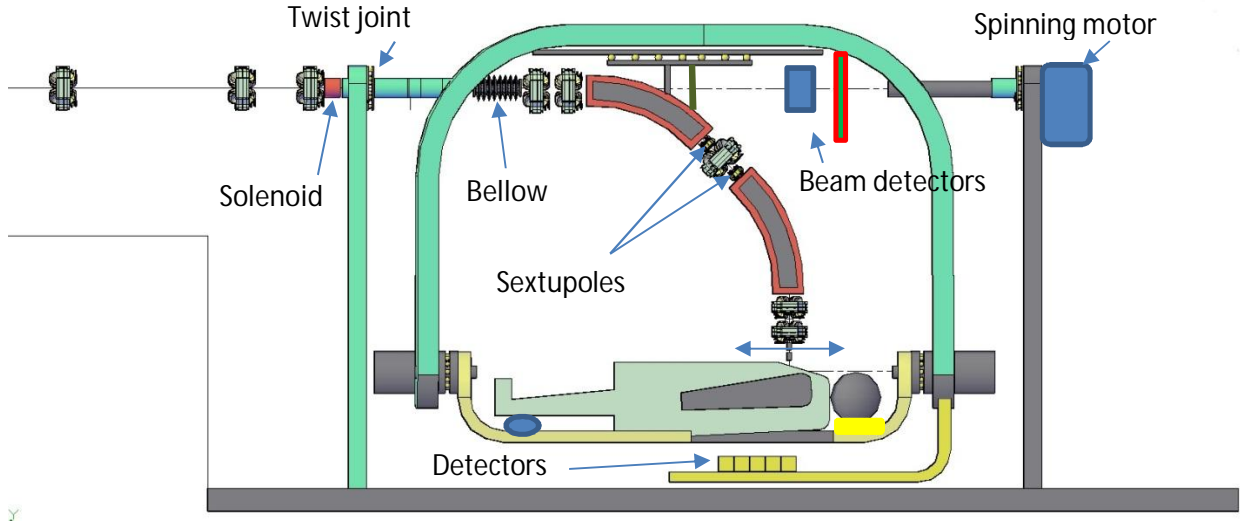

**Figure 12.** Gantry system zoomed. Bellow allows longitudinal motion of all elements located on gantry within ±10cm. The frame which supports the optical elements on gantry is not shown here.

The lodgment for the patient is long enough, so the focal point could be arranged for any part of the body. In addition to that, the magnets and lenses located on the gantry have the freedom to be moved longitudinally together as the envelope function is horizontal here along the beam line down to the first quadrupole lens, $b\phi(s) = 2s / b_0$, $b\phi_{x,y}(s) < 0.125$ for the distance $s = \pm 10$ $cm$ (this option marked by double-headed arrow in Fig.12). Beam detectors at continuation of straight channel serve for tuning the proton/ion beam, while preparing it for service.

Some detectors located below the beam trajectory crossing the patient, are able to register the charged particles passed through the patient. They are attached to the gantry also (marked as Detectors in Fig.12). These detectors represent a pixel calorimeter which is able to restore the 3D profile of tumor during tomography with the low-intensity beam. Positioning of these detectors correlates with the longitudinal position of elements of optical channel in the gantry.

The beam optical channel is attached to the gantry with the help of linear bearings, seen at the top of Fig.12.

Magnetic characteristics of this channel remains within technical limitations, say the current density in coils are below 10 $A/cm^2$, which requires a water cooling conductor, however. For a 300 $MeV$ proton beam, the distance before stop in water is ~100cm, so the tomography is possible here with low intensity beam.

Magnetic field value in a bending magnet with the bending radius ~40 $cm$ comes to be H~20kG, which requires the value of Ampere-turns $N > I[A] = H[G] > a[cm] / 0.4p$, where $a$ stands for the magnet aperture. As the beam size is small, ~2 $mm$ only (Fig.10) then the aperture could be made ~5$mm$, so the number of ampere-turns comes to $N > I[A] = 80kA$. For the current density 20$A/mm^2$ that will require the area of conductor cross section ~4000 $mm^2$ i.e.~65x65 $mm$. These numbers are indicative; exact values must be recalculated for exact choice of maximal energy of the beam. Also, the bending radius could be increased, as now in our scheme the transverse dimension of active bend comes to be about the bending radius of the magnet (with some extensions for the quad and sextupoles) so the 1-1.5 *m* total transverse distance looks feasible. This gives an idea of how big the radius of circle in Fig.3 is.

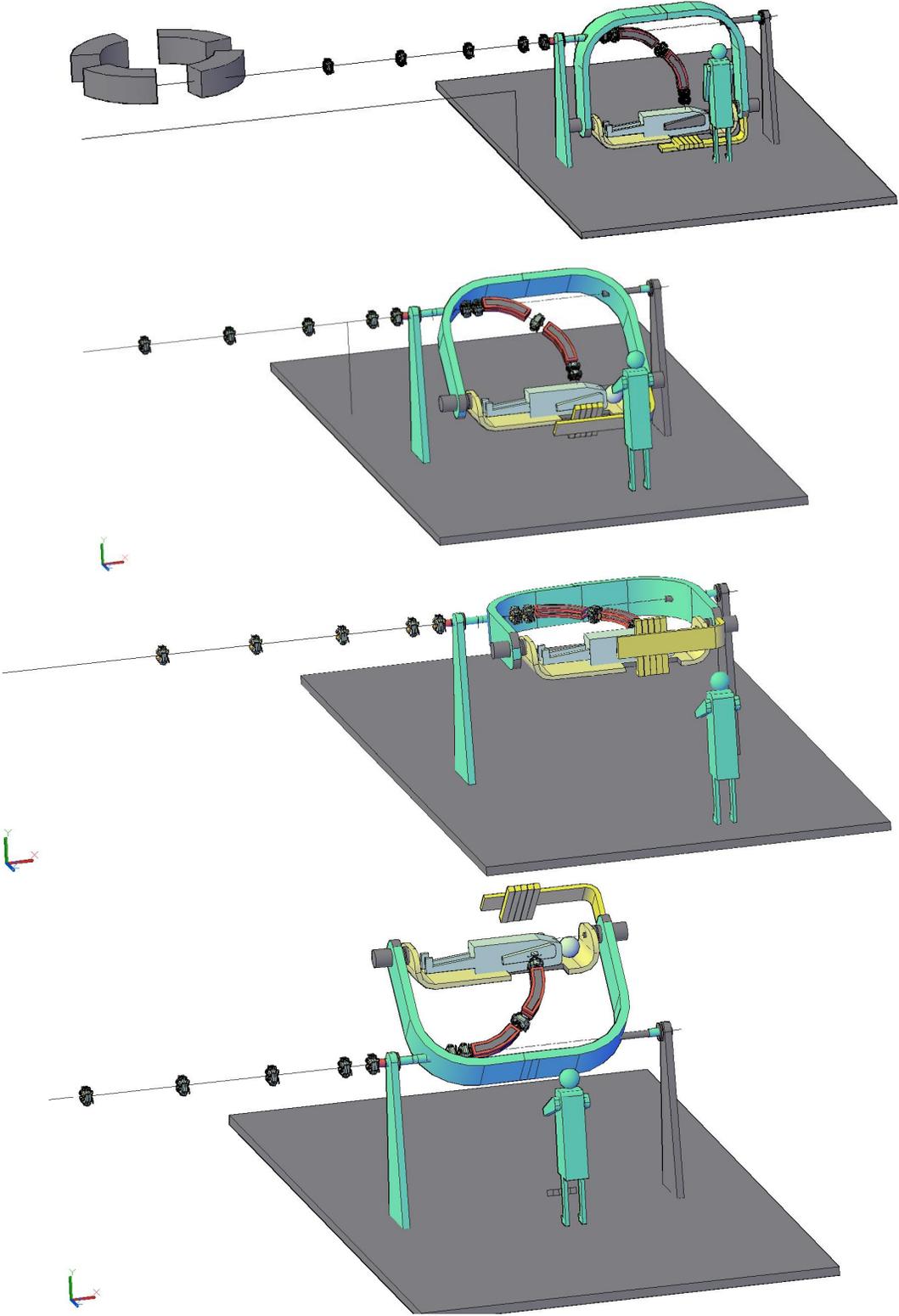

**Figure 13.** Four stages of rotation represented in isometric projection.

## SUMMARY OF PROPOSAL

An object of this proposal is a compact system for the irradiation of a patient in a comfortable position with reduced irradiation of tissues neighboring to the tumor.

The current proposal offers the following:

a) More compact gantry system. Practically, the number of magnetic elements is half of what holds the traditional scheme;

b) Comfortable relaxed position for the patient, who is lying horizontally in cradle which is linked to the gantry;

c) Additional reduction of dosage for neighboring regions of tumor due to quick enlargement of the beam size before (and right after) the focal point;

d) Possibility for the optics to adjust the focal point location in three coordinates, which is useful for making the distribution of dosage over the object (tumor) more precise; it is also useful while doing tomography of tumor with this equipment.

e) Additional registration system located from the other side behind the patient at all times allows control of irradiation *in situ*.

f) Possibility to move the focal point along the patient simply by mechanical motion of elements located on the gantry as a whole (trombone). This mode is allowable as the envelope functions have small derivative $b\phi'_{x,y}(s) @ 0$ at the place where the gantry linked to the stationary channel.

g) Usage of electron cooling techniques could provide even smaller beam size for the proton/ion beams.

## CONCLUSION

By attachment of the patient's body carrier (lodgment) to the gantry, the new possibility of arrangement of irradiation of a tumor under different angles becomes open.

The new system comes out to be more compact, less expensive and more reliable. In addition to this improvement in hardware, the optical system associated with this new method allows having the waist of beam inside the tumor. If the focal distance of final doublet made short enough, then the beam envelope is strongly divergent from the focal point (hourglass effect).

The same concept could be used with X-ray beam, although this might be useful for irradiation of tissues on the surface of body mainly, as advantages of Bragg's pick of energy deposition is not present for X-ray source.

We hope that this method will be useful in a noble activity of accelerator physicists on a field of medicine.